\documentstyle[12pt,fleqn]{article}
\def\singlespace {\smallskipamount=3.75pt plus1pt minus1pt
                  \medskipamount=7.5pt plus2pt minus2pt
                  \bigskipamount=15pt plus4pt minus4pt
                  \normalbaselineskip=15pt plus0pt minus0pt
                  \normallineskip=1pt
                  \normallineskiplimit=0pt
                  \jot=3.75pt
                  {\def\smallskip {\vskip\smallskipamount}}
                  {\def\medskip   {\vskip\medskipamount}}
                  {\def\bigskip   {\vskip\bigskipamount}}
                  {\setbox\strutbox=\hbox{\vrule 
                    height10.5pt depth4.5pt width 0pt}}
                  \parskip 7.5pt
                  \normalbaselines}

\begin{document}

\textheight 9.0in \topmargin -0.5in \textwidth 6.5in \oddsidemargin -0.01in

\bigskip

\begin{center}
{\bf {\Large Quantum Mechanics without Spacetime}}

\smallskip

{\bf {\ {\it - A Possible Case for Noncommutative Differential Geometry? -} }}
\end{center}

\medskip
\centerline{\bf T. P. Singh}
\smallskip
\centerline{\small {\it Tata Institute of Fundamental Research, Homi 
Bhabha Road, Mumbai 400 005, India}} 
\centerline{{\small \tt email address: tpsingh@tifr.res.in}}

\bigskip

\centerline{\bf Abstract}

{\small The rules of quantum mechanics require a time coordinate for their
formulation. However, a notion of time is in general possible only when a
classical spacetime geometry exists. Such a geometry is itself produced by
classical matter sources. Thus it could be said that the currently known 
formulation of quantum mechanics  pre-assumes the presence of classical 
matter fields. A more fundamental formulation of quantum mechanics should
exist, which avoids having to use a notion of time. In this paper we discuss
as to how such a fundamental formulation could be constructed
for single particle, non-relativistic quantum mechanics. We argue that 
there is an underlying non-linear theory of quantum gravity, to which 
both standard quantum mechanics and classical general relativity are 
approximations. The timeless formulation of quantum mechanics follows from the 
underlying theory when the mass of the particle 
is much smaller than Planck mass. On the other hand, when the particle's
mass is much larger than Planck mass, spacetime emerges and the underlying 
theory should reduce to classical mechanics and general relativity. We also 
suggest that noncommutative differential geometry is a possible candidate 
for describing this underlying theory.}      

\bigskip

\singlespace

The rules of quantum mechanics require the concept of time, for their
formulation. The time coordinate determines the choice of canonical position
and momenta, the normalization of the wavefunction, and of course, the
evolution of the quantum system. From the point of view of the special
theory of relativity, time is a component of spacetime; furthermore the
general theory of relativity endows the spacetime with a pseudo-Riemannean
geometry. This geometry however, is determined by the distribution of {\it %
classical} matter - such matter is of course a limiting case of matter
obeying the rules of quantum mechanics. Hence, via its dependence on time,
quantum mechanics pre-assumes the existence of classical matter, whose very
properties it should explain in the first place. A more fundamental
formulation of quantum mechanics should exist, which does not refer to a
background time. (for a recent discussion on the difficulties associated
with quantum clocks see e.g. \cite{Carlip}). The purpose of the present paper 
is to outline a proposal for such a fundamental formulation, which we call 
Fundamental Quantum Mechanics (FQM).

The need for FQM also follows from noting that the Universe could, in
principle, be completely devoid of classical matter. For instance, the
Universe could consist entirely of non-relativistic, non-classical
microscopic particles, which would be described by standard non-relativistic
quantum mechanics, if a background spacetime were to be available. Since
here classical matter and consequently classical spacetime geometry are
completely absent, FQM is necessary for describing this Universe. Under very
special circumstances (suitably chosen quantum states etc.) such a Universe
could be described by a semiclassical theory of gravity (classical gravity
produced by quantum matter), but the semiclassical description is in general
not valid (for a useful discussion see \cite{FordKuo}). We observe that FQM
can become necessary even if the typical energy scale of the particles in
the system is much less than the Planck scale, and the particles are `moving'
non-relativistically.

In order to attempt tackling the simpler problem first, in this paper we
address only the issue of a FQM for non-relativistic quantum mechanics. Even
though its true that here time and space are absolute Newtonian concepts,
they are nonetheless attributes of a classical world, and a nonrelativistic
approximation to the geometric description provided by the special and
general theories of relativity. It is therefore meaningful to search for an
FQM corresponding to non-relativistic quantum mechanics.
(for another discussion on quantum mechanics without time, see
\cite{Rovelli}).

We can learn about the properties of FQM by constructing a simple model of
the Universe. Imagine, for the sake of visualization, the Universe to be a
manifold which is a 2-sphere, with one angular coordinate representing
space, and the other time (unphysical properties like closed timelike curves
are not relevant here since the spherical topology is only chosen as a model
which makes it easy to form pictures). Next, let us imagine that there is
only one object in this Universe - a macroscopic, localized object of mass $%
m_1$. If $m_1$ can be treated classically, then it appears reasonable to
propose that the spacetime on this manifold, as well its accompanying
pseudo-Riemannean geometry, are due to this object. The object's own state
is of course a trajectory in this spacetime, or equivalently, in the phase
space. A key feature of this classical state is the absence of
superpositions.

Next, we consider the case that the mass $m_1$ is so small that the object is no
longer classical, but must be treated according to quantum mechanics.This
would be straightforward if there were a background spacetime on the
2-sphere. Now, however, we have a completely delocalized particle which can
be thought of as residing on the 2-sphere, but there is no longer any
Riemannean spacetime geometry on the manifold, nor any concepts of space and
time. The particle, manifold and its pseudo-Riemannean geometry have all
merged into one, in some sense. The dynamics of the system can be described,
possibly, as a geometry of the 2-sphere - we call this dynamics
Nonrelativistic Quantum Gravity (NQG). The state of the system can no longer
be said to have a causal evolution, but exists as a whole, once and for all.

Another way to arrive at this conclusion is to consider the situation where
the 2-sphere already possesses a classical spacetime geometry because of
existent classical matter, and the value of the mass $m_1$ (now assumed to
be a test particle) is gradually reduced. When $m_1$ is classical, its state
is a spacetime trajectory, but when it becomes quantum mechanical, the state
becomes an element of Hilbert space, labeled by a time coordinate. It is
rather unnatural to imagine that as the value of $m_1$ is reduced, the state
at some stage jumps from being in physical spacetime, to being in a Hilbert
space. It is more natural to assume that the state of $m_1$ always belongs
to the fundamental NQG 2-sphere - which looks like physical spacetime for
large values of $m_1$ and like the direct product of a Hilbert space with
time, for small values of $m_1$.

The only natural scale which separates a large classical value for the mass $%
m_1$from a small `quantum mechanical' value should be determined by Planck
mass $m_{Pl}=(\hbar c/G)^{1/2}\sim 10^{-5}$ gm. Now one expects, from
observation, the mass scale separating the `classical' from the `quantum' to
be smaller than $m_{Pl}$ by a few orders of magnitude. However, for want of
a better understanding, we will for the time being continue to refer to this
separation scale as $m_{Pl}$, with the understanding that a more refined
analysis will yield a somewhat smaller separation scale.  For $m_1\gg m_{Pl}$
(equivalently $\hbar\rightarrow 0$ or $G\rightarrow \infty $) NQG should reduce
to a classical spacetime trajectory for the particle, and to the
nonrelativistic Einstein equations (i.e. Newtonian gravity) for the emergent
spacetime geometry. Since classical objects are never observed in superposed
states, it has to be the case that NQG is a non-linear theory - two
solutions of the theory cannot be superposed.

It is important to note that NQG is {\it not} the timeless description of
non-relativistic quantum mechanics (i.e. FQM) that we are seeking. Unlike
standard quantum mechanics and unlike FQM, NQG is non-linear, and involves
the gravitational constant $G,$ since it must also describe the
gravitational effects of $m_1$. It is clear that if we take the limit $%
G\rightarrow 0$ (equivalently, $m_{Pl}\rightarrow \infty $ or $m_1\ll m_{Pl}$%
) then NQG no longer describes the gravity of $m_1,$ and should reduce to
FQM: a linear theory which is the timeless formulation of standard
non-relativistic quantum mechanics. In FQM too, the physical state of $m_1$
does not have a causal evolution, but exists as a timeless whole on the
2-sphere.

We thus have the picture that the dynamics of the object $m_1$ on the
2-sphere is in general described by NQG - here spacetime geometry and matter
cannot be separated from each other. NQG reduces on the one hand (when $%
m_1\gg m_{Pl}$) to non-relativistic Einstein equations and classical
mechanics (the coordinates of the sphere now become space and time), and on
the other hand (when $m_1\ll m_{Pl})$ to a timeless description of standard
quantum mechanics.

The time-dependent equivalent version of the FQM for $m_1$ is obtained when
the 2-sphere possesses a classical spacetime geometry due to other matter
sources. This could happen, for instance, if there is present on the
2-sphere, another mass $m_2\gg m_{Pl}$ which hence endows it with a
classical spacetime. $m_1$ is now a test particle on this 2-sphere in the
sense that $m_1\ll m_{Pl}\ll m_2$. The FQM for $m_1$ now has the standard
interpretation of non-relativistic quantum mechanics. It could be concluded
from this discussion that there is an apparent time evolution in quantum
mechanics only because the Universe today is dominated by classical matter,
which produces classical spacetime. At a deeper level, quantum mechanics
describes the physical state of the system $m_1$ not as an evolution in
time, but as a dynamics in which spacetime and matter cannot be separated.

Important clues as to the nature of NQG can be obtained by examining the
possible gravitational interaction a test particle $m_1$ has with the
background gravitational field provided by the particle $m_2$. The various
possibilities are shown in the Table below, depending on how $m_1$ and $m_2$
compare with $m_{Pl}$.

\begin{center}
$
\begin{array}{ccccc}
& m_2\rightarrow : & m_2\ll m_{Pl}, & m_2\sim m_{Pl}, & m_2\gg m_{Pl} \\ 
m_1\downarrow &  &  &  &  \\ 
m_1\ll m_{Pl} &  & NO & Test1 & NRQM \\ 
m_1\sim m_{Pl} &  & X & NQG & Test2 \\ 
m_1\gg m_{Pl} &  & X & X & GR 
\end{array}
$
\end{center}

The $11$ entry (NO) indicates that gravitational interaction is absent when
both the masses are much smaller than $m_{Pl}$. The 2$1$ and 31 entries (X)
indicate that in these cases $m_1$ is not a test particle. The $12$ entry
(Test1) is the most important, as this is the domain where laboratory
experiments could in principle search for possible signatures of NQG , by
studying the gravitational interaction of a quantum mechanical particle ($%
m_1 $) with the NQG `gravitational field' of $m_2$. The $22$ entry (NQG)
represents the genuine gravitational interaction of two particles in NQG:
this would describe the dynamics that replaces Newtonian gravity in a
quantum theory of gravity. The $13$ (NRQM) entry is the standard
non-relativistic quantum mechanics on a background spacetime, whereas the $%
23 $ entry (Test2) probes the response of a particle in NQG to a classical
spacetime geometry. The final $33$ entry is Newtonian gravity
(nonrelativistic GR). Looking at the last column, since both the first
(NRQM) and the third (GR) entry satisfy the equivalence principle, one could
expect that NQG on a background spacetime (middle entry Test2) satisfies the
equivalence principle too - a possible indicator that NQG is a generally
covariant theory.

We have been able to infer some properties of nonrelativistic quantum
gravity by starting from the premise that there should be a description of
quantum mechanics which does not refer to a time coordinate. The theory to
which one is led (NQG) has as its limits both quantum mechanics ($%
G\rightarrow 0$) as well as nonrelativistic gravity ($\hbar\rightarrow 0$). It
is hence possible that the dynamics of NQG (in our model the dynamics of the
2-sphere) is described by a noncommutative differential geometry (NDG)
\cite{Connes},
because NDG has within itself, as special cases, a commutative spacetime
geometry which describes spacetime and gravity, and also, a noncommutative
algebra structure which can describe quantum mechanics on an ordinary
spacetime.

An outline of the dynamics of NQG on the 2-sphere is suggested here. We
shall assume that the 2-sphere is a noncommutative space on which the
algebra of functions is in general noncommuting. Let $A$ and $B$ denote the
`coordinates' on the 2-sphere, and let there be, associated with these
`coordinates', `momenta' $p_A$ and $p_B.$ These four quantities are assumed
to describe a particle $m$ on the 2-sphere. We propose, on the 2-sphere, 
commutation relations of the following kind:

$$
[A,B]=iL_{Pl}^{2} F_1(m/m_{Pl}) 
$$
$$
[A,p_A]=[B,p_B]=i\hbar F_2(m/m_{Pl}) 
$$
$$
[p_{A},p_B]=i{\hbar^{2}\over L_{Pl}^{2}} F_3(m/m_{Pl}) 
$$

The functions $F_1$, $F_2$ and $F_3$ are assumed to go to zero in the limit $m\gg
m_{Pl}$. Thus all the four quantities describing the particle become commuting in
the large mass limit. Now one may identify $A$ and $B$ with ordinary spacetime
coordinates on the 2-sphere, and $p_{A}$, $p_{B}$ with ordinary energy and momentum.

In the limit $m\ll m_{Pl}$ these commutation relations describe FQM. Since
a background spacetime is absent, the function $F_{1}(m/m_{Pl})$ has to
be non-vanishing in this limit. FQM is hence described in terms of quantities
which are not standard spacetime coordinates and momenta. This description 
becomes equivalent to standard quantum mechanics in the presence of an
external spacetime, though it is unclear at present as to how that
happens. The presence of the mass $m$ in the commutation relation [A,B] should
be seen as analogous to the fact that in Riemannean geometry the non-commutativity 
of covariant derivatives is determined by the Riemann tensor, which itself
is related to the distribution of matter. Hence we are suggesting that at
a fundamental level not only the covariant derivatives, but the coordinates
as well, do not commute, and the non-commutativity of the latter may also
be related to matter.

The physical state of the system in NQG is the noncommutative analog of a
vector field, or equivalently a noncommutative analog of a derivation, on
the 2-sphere. Furthermore, associated with the differential structure of the 
$A,B$ space there is a concept of curvature of the noncommutative space. The
exact implementation of the notion of curvature in NDG is at present an
issue that has not been fully resolved \cite{Madore}.
 We propose that this curvature is induced by the
presence of the mass $m$, in the spirit of general relativity. In the
commutative limit the dynamical equation relating curvature to $m$ reduces
to Einstein's equations. The detailed nature of the dynamics of NQG in the
language of noncommutative geometry is under investigation. 

It appears to us that, like spacetime, the metric is also an emergent
concept, valid only when the Universe is dominated by classical matter. Our
non-relativistic considerations here cannot explain the Lorentzian signature
for the emergent metric. We hope to address relativistic generalizations of
these ideas in the near future. 

Our proposal for the conceptual structure of NQG has, built within itself,
the attractive possibility of addressing the issues of wave function
collapse during a quantum measurement, and the EPR paradox. From our
viewpoint, during a quantum measurement, there is a sudden jump in the value
of $m_1$ from the range  $m\ll m_{Pl}$ to the range $m\gg m_{Pl}$ (as a
microscopic quantum system comes into contact with a macroscopic measuring
apparatus). The associated geometry changes from the delocalized, timeless
description of the 2-sphere on the NQG, to a classical spacetime trajectory
- this could help understand the apparently acausal change associated with
an EPR measurement. Also, a proper understanding of NQG could explain the
origin of the $\mid \psi \mid ^2$ probability rule which comes into play
during a quantum measurement. The apparent breakdown of the unitarity of the
Schrodinger equation during a measurement is possibly associated with the
fact that NQG makes a transition from being a linear theory for $m\ll m_{Pl}$
to being the non-linear theory which it actually is, when $m\gg m_{Pl}$. Our
picture also supports Penrose's idea that gravity is involved in the
collapse of the wave function \cite{Penrose}.

It is also of interest to investigate what possible connection this
work might have with string theory and M-theory, wherein the position
coordinates of a D-brane become non-commuting.

We conclude by suggesting that Einstein's criticism that quantum mechanics
is incomplete may now be understood as the theory having to refer to a
background time. A version of the theory which does not refer to such a
background time possibly removes this incompleteness.

\bigskip

\centerline{\bf ACKNOWLEDGEMENTS}

\noindent I thank Sukratu Barve, Pankaj Joshi and Cenalo Vaz for useful 
discussions. I acknowledge the partial support of FCT, Portugal under 
contract number SAPIENS/32694/99.

\end{document}